\documentclass{pazhbabel}

\usepackage{natbib}
\usepackage{graphicx}
\usepackage{amsmath}
\usepackage{epsfig,graphics}
\usepackage{rotating} 
\usepackage{epsfig}

\newcommand {\be}{\begin{equation}}
\newcommand {\ee}{\end{equation}}

\begin{document}

\title{Estimation of Plasma Parameters in an Accretion Column
near the Surface of Accreting White Dwarfs from Their Flux Variability}
\author{Andrey N. Semena\email{san@iki.rssi.ru}\address{1}, Mikhail G. Revnivtsev\address{1},
\addresstext{1}{Space Research Institute, Russian Academy of Sciences, Profsoyuznaya 84/32, 117997 Moscow, Russia\\}}
\journalinfo{2012}{38}{5}{364}




\thispagestyle{empty}
\begin{abstract}
We consider the behavior of matter in the accretion column that emerges under accretion in binary systems near the surface of a white dwarf. The plasma heated in a standing shock wave near the white dwarf surface efficiently radiates in the X-ray energy band. We suggest a method for estimating post-shock plasma parameters, such as the density, temperature, and height of the hot zone, from the power spectrum of its X-ray luminosity  variability. The method is based on the fact that the flux variability amplitude for the hot region at various Fourier frequencies depends significantly on its cooling time, which is determined by the parameters of the hot zone in the accretion column. This allows the density and  temperature of the hot matter to be estimated. We show that the characteristic cooling time can be efficiently determined from the break frequency in the power spectrum of the X-ray flux variability for accreting white dwarfs. The currently available X-ray instruments do not allow such measurements to be made because of an insufficient collecting area, but this will most likely become possible with new-generation large-area X-ray spectrometers.
\medskip
\end{abstract}

\section*{Introduction}
The accretion of matter onto a white dwarf with a magnetic field is accompanied by the formation of a flow channeled onto its magnetic poles. As the surface is approached, the matter moving along the magnetic field lines is a highly supersonic flow that decelerates sharply and heats up in a  standing shock  wave. Thus, a hot (with a temperature of 5 - 30 keV)  zone is formed between the white dwarf surface and the shock. The hot zone is a powerful source of X-ray emission and, if the cyclotron losses in the overall energy balance of the hot region are important, optical/infrared  emission \citep{aizu73,fabian76}.

\cite{langer81} considered the behavior of the hot zone in an accretion column by assuming a constant accretion-channel cross section, the absence of gravity, and a cooling law  $dE/dt=\Lambda \sim \rho^2 T^\alpha$. A global thermal instability of the hot zone resulting in quasi-periodic oscillations of its height and total luminosity was shown to take place in such a system at $\alpha < 1.6$. Subsequently, it was shown in more accurate calculations that the values of $\alpha$ at which the flow is stabilized ($\alpha_{st} \sim 0.6 - 1.2$), are generally lower than those obtained by  \cite{langer81} and depend on the boundary conditions \citep{mignone05}. The oscillation period for such a system is of the order the characteristic energy loss time in the hot zone of the accretion column. The height of the hot zone is determined by the accretion rate, the column cross section, and the cooling law. Subsequently,  \cite{langer82} performed one-dimensional numerical simulations of an accretion column with a hot zone that confirmed the main conclusions of their previous paper. Later on, \cite{imamura96}carried out a perturbative analysis of the system. As
a result, oscillations in the height of the hot zone at  frequencies of $\sim 1$ Hz.
	
The global thermal instability of the hot region is a convenient observational manifestation of accretion onto magnetized white dwarfs that could be used to estimate the fundamental parameters of the accretion flow in binary systems. In some binary systems, the oscillations in the height of the hot zone may have been detected (see, e.g., the review by   \citealt{larsson95}),  but by no means in all magnetic systems (see, e.g.,\citealt{imamura00,drake09}),suggesting that the oscillation emergence problem has not been solved. In any case, observations show that the luminosity of white dwarfs often has no distinct quasi-periodic oscillations. One would think that this does not allow the mechanism of thermal instabilities to be used to estimate the parameters of the matter in an accretion column. 

Here, we show that the properties of the hot zone can be estimated not only from the quasi-periodic oscillation frequency but also from the shape of the broadband power spectrum for the light curve of accreting magnetized white dwarfs. The method is based on the fact that the accretion-rate modulations always present in an accretion flow in a wide range of Fourier frequencies cannot be directly translated into accretion-column luminosity modulations, because the energy release in the accretion column occurs in a finite time determined by the global parameters of the matter in the column - density, the mass flux in the column, etc. Thus, the power spectrum of the accretion-column luminosity variability must contain a feature, a break, at frequencies of the order of the reciprocal matter cooling time in the hot region. We demonstrate this effect through numerical simulations.

\section*{Thermal instability of the hot post-shock region in accretion column}
	
\begin{figure*}
\hfill
\includegraphics[width=\textwidth]{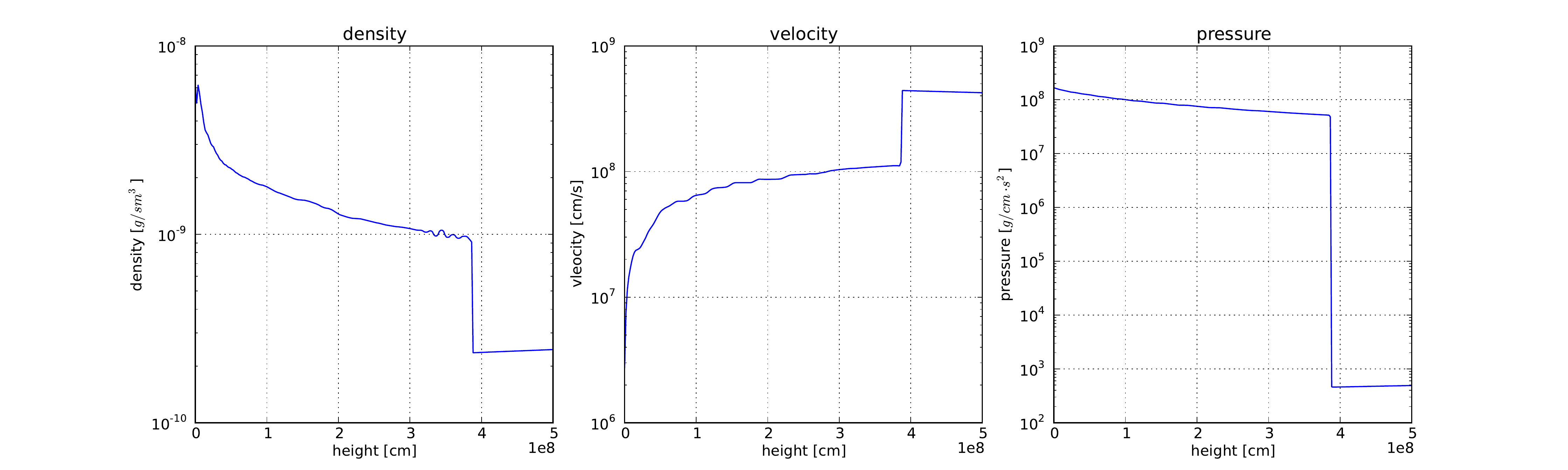}
\caption{\small Density, pressure and velocity profiles along accretion column. Accretion rate in presented calculation is $10^{-12} M_\odot$/year,  cooling rate $\Lambda = \Lambda_{bremss}$, white dwarf mass M = $M_\odot$, accretion column cross section area A = $10^{15}$ cm$^2$.}
\label{profiles}
\end{figure*}

The appearance of the oscillations detected by \cite{langer81,langer82}, can be explained as follows:
as the shock moves away from the white dwarf surface, the gas passing through the shock has a higher temperature than in the case of a standing wave (because it strikes the shock surface with a velocity $\dot{h} + v_{ff} > v_{ff}$ where $\dot{h}$ is the shock front velocity), post-shock pressure increases in this case. As the shock moves upward, the total energy losses from the hot zone grow through an increase in the cooling path length. In the course of time, this leads to a drop in post-shock pressure, its stopping, and the backward motion of the shock toward the white dwarf surface. As the shock moves toward the white dwarf surface, the reverse situation arises?the matter passing through the shock has a lower temperature and the pressure drops, but, at the same time, the total losses also drop. Thus, other things being equal, the oscillations of the hot zone are determined by the matter cooling function (different oscillation periods can be obtained for the same ratio of the accretion rate to the  accretion-channel cross section $\dot{M}/A$). It is worth noting that for the cooling law in the case of a simple optically thin plasma (bremsstrahlung) $\Lambda_{\rm bremss}\propto \rho^2 T^{0.5}$ the cooling efficiency increases rapidly with matter density and the maximum accretion-column height at high accretion rates is lower than that at lower ones; the oscillation period decreases simultaneously with height.

The characteristic time of such oscillations $\tau$ is related to the cooling time in the accretion column, which can be roughly estimated as the ratio of the post-shock energy density $E_{ps}$ to the cooling rate $\Lambda_{ps}$.

\begin{equation}
\tau \approx \frac{E_{ps}}{\Lambda_{ps}} = \frac{k_b \rho_{ps} T_{ps}}{(\gamma - 1) \mu m_p \Lambda_{ps}}
\label{t_cooling}
\end{equation}

In the considered problem, we have a strong shock in which the change in post-shock density $\rho_{ps} = 4 \rho_{ff}$ ($\rho_{ff} = \dot{M}/A v_{ff} = \dot{M}/A\sqrt{R/2GM}$). Where $\rho_{ff}$ -- is the pre-shock matter density, A is the cross-sectional area of the accretion channel, M and R are the white dwarf mass and radius, $\gamma$ is the adiabatic index, $\mu$ is the molecular weight, and kb is the Boltzmann constant. In this case, the post-shock temperature is
$$
T_{ps} \approx \frac{3}{16}\frac{\mu m_p}{k_b} v_{ff}^2
$$

 In case of energy loss function $\Lambda_{\rm bremss}$ ($\approx 6 \cdot 10^{20} \rho^2 \sqrt{T}$ [erg/cm$^{3}$/s], \citealt{rybicki79}) the characteristic time is
$$
\tau \approx 1.2 \cdot 10^{-24} \frac{A}{\dot{M}} \frac{M}{R} \approx 2~ \textrm{сек}
$$

for $M = M_\odot; A = 10^{15}$ cm$^2$; $\dot{M} = 10^{15}$g/sec$\approx1.6\times10^{-11} M_\odot/$year; $R =  10^9$ cm. Height of the host zone can be estimated by relation $h \sim v_{sh}\tau$ (\cite{langer81}).

In the more realistic problem of a matter flow in an accretion column, one should consider a number of additions that can affect significantly the general behavior of the hot zone. The main additions include the possibility of describing a two-temperature electron-ion flow in the hot region, the introduction of a more complex energy loss law with cyclotron losses, and various boundary conditions at the bottom of the accretion column. A two-temperature flow was considered by \cite{imamura96,wu99, saxton99}. Two-temperature flows were shown to be less stable than one-temperature ones with respect to a change in the coefficient $\alpha$ for a $\rho^2 T^\alpha$ cooling low. Problems with various realizations of boundary conditions were considered by Saxton and Wu \cite{saxton01, saxton02} and \cite{mignone05}. These authors showed that variations of boundary conditions with a sink affect weakly the behavior of the hot zone.

\begin{figure}[t]
\centering
\includegraphics[width=1.1\columnwidth]{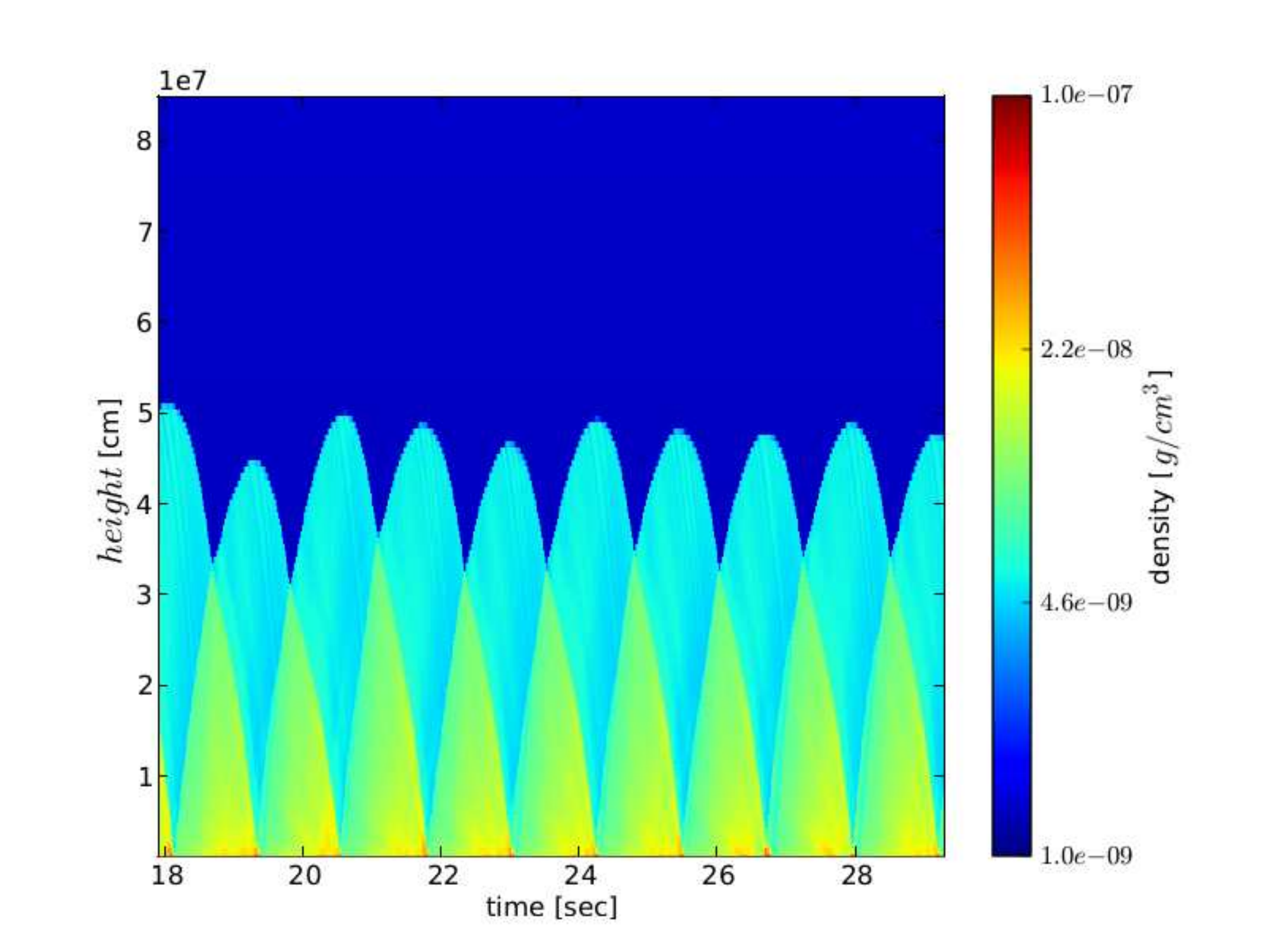}
\caption{\small
Density profile time dependence along accretion column. QPO is clearly visible. Accretion rate in computation is $\dot{M} = 10^{-11} M_\odot/$year, other parameter is equal to one from pic. \ref{profiles}}
\label{qpos}
\end{figure}

	\cite{cls85} showed that introducing the additional term responsible for the cyclotron losses into the cooling function causes the oscillations to be suppressed. Subsequently, the problem was also considered by \cite{saxton97,saxton98}. In the above papers, a cyclotron loss function of the following form was used  (\cite{langer82} \cite{Wu94}): 
\begin{equation}
\Lambda_{\rm cyc} \propto \rho^{0.15} T^{2.5}
\label{cyc}
\end{equation} 
	 This cooling function was derived under the following assumptions:
 \cite{chanmugam79} showed that self-absorption is inherent in cyclotron radiation up to frequencies $\omega^* =m^* \omega_c$($m^*\sim$10). Subsequently \cite{wada80} give a simple approximation for $m*$: $m* = 9.87 \cdot 10^{-4} T^{0.5} (4 \pi n_e e x/(B \cdot 10^7))^{0.05} $ where B - magnetic field, x - s the distance from the emitting layer to the shock, $n_e$ - electron number density, T is the post-shock temperature. \cite{langer82}suggested the following solution for the cyclotron losses of the column: a layer at depth x relative to the shock front with thickness dx will radiate due to cyclotron emission the blackbody flux difference  $L_{bb}(\omega^*(x + dx)) - L_{bb}(\omega^*(x))$, which corresponds to cyclotron frequencies ($\omega > \omega^*$).For the cyclotron losses at these frequencies, the column is optically thin. They also showed that the column is optically thin for thermal losses corresponding to the plasma temperature in the column hot zone. In the expression derived by \cite{langer82} $\Lambda_{cyc} \propto \rho^{0.15} T^{2.5} x^{-0.85}$ (where x is the distance from the emitting zone to the shock). In recent papers (see, e.g., \cite{saxton99})  a simplified loss function with losses independent of the emitting-layer depth (\ref{cyc}) was used. In our paper, the main objective was to model the hot zone with suppressed oscillations. Since the loss function (\ref{cyc}) allows the oscillations of the hot zone to be suppressed and corresponds to the general form of cyclotron losses, we decided to use it in our model.
	
	The suppression of the oscillations can be explained by the fact that the cyclotron losses dominate in the immediate neighborhood of the shock, where the matter temperature in the hot zone is highest. As a result, the cyclotron losses depend weakly on the column height. In addition, the cyclotron losses are in phase with the shock motion: as the shock moves toward the white dwarf surface, the cyclotron losses decrease together with the post-shock matter temperature and increase as the shock moves in the opposite direction. Thus, the cyclotron radiation suppresses the instability development mechanism. It is worth noting that the cooling through bremsstrahlung $\Lambda_{\rm bremss}$ increases with accretion rate faster than the cyclotron cooling $\Lambda_{\rm cyc}$ and, therefore, a stronger magnetic field is required for a system with a higher accretion rate to suppress the oscillations.

In real accretion flows, the accretion rate often exhibits aperiodic variability - flickering (see, e.g., \citealt{bruch92}). As a rule, the power spectrum of the accretion rate variability is described by a power law of the form $P(f)\propto f^{-1...-2}$. In the case of accretion
onto magnetized neutron stars, the accretion rate through magnetospheric flows varied at frequencies up to hundreds of Hz (see, e.g., \cite{jernigan00}).  The effect of noise on the instability was considered by \cite{wolff91}, who showed that the hot zone becomes more unstable at a noisy accretion rate. They also pointed out that the characteristic oscillation time changes when passing from low accretion rates to high ones as a result of the change in the relationship between the cyclotron and thermal losses. In this case, the oscillations are still suppressed when the cyclotron losses dominate.	


\section*{NUMERICAL SIMULATIONS OF AN ACCRETION COLUMN}	 

\begin{figure}
\centering 
\includegraphics[width=1.1\linewidth, height = 250pt]{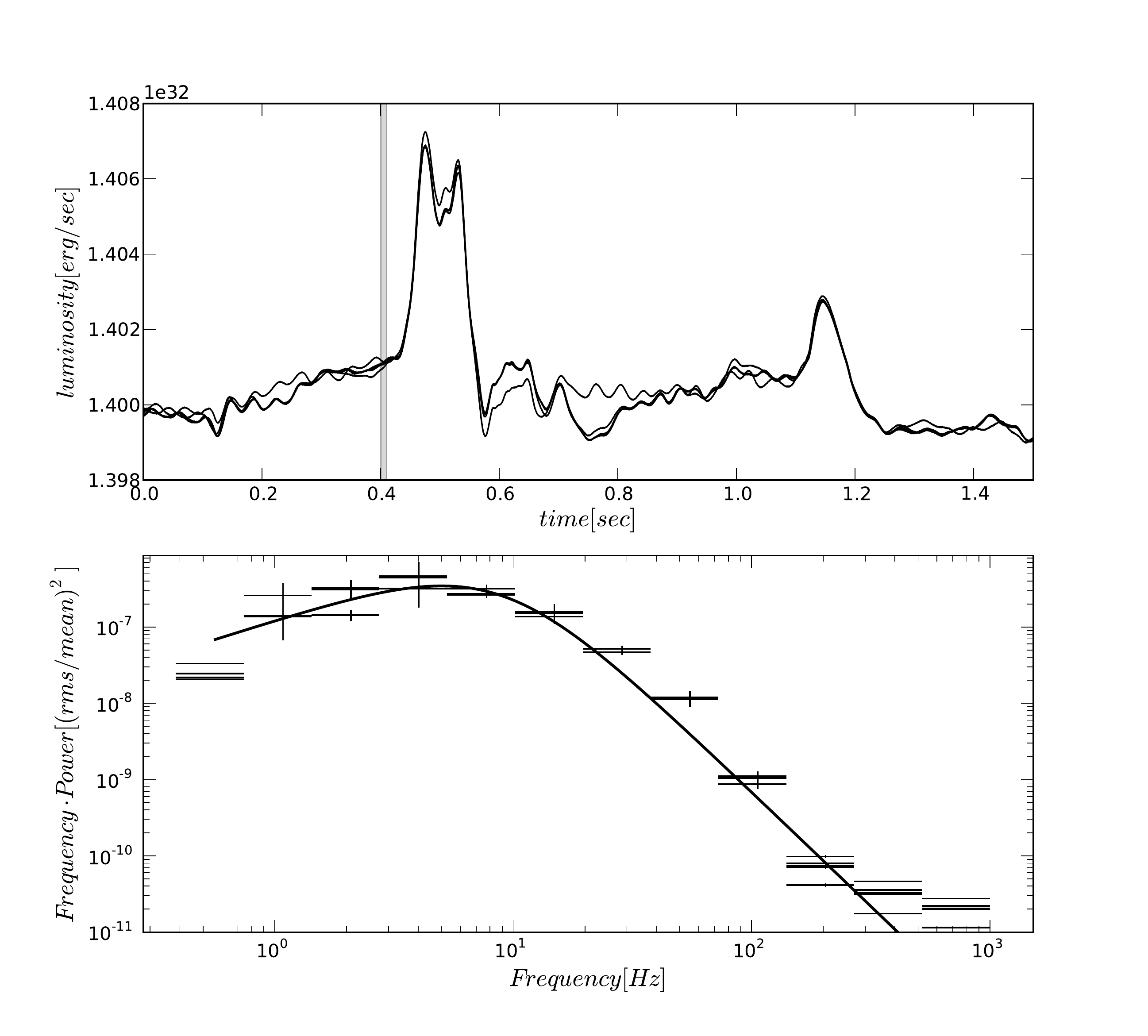}
\caption{
\small
Lignt curve responce on short -0.01 sec 5\% amplitude perturbation of accreiton rate (accretion rate $10^{-11} M_\odot/$year). In top field plotted five signals, grey vertical line denotes accretion rate 5\% excess duration. In bottom field power spectrums of this signals plotted. Fitting by function with break give break frequency $\sim$12 Hz.}
\label{delta}
\end{figure}

Our objective was to show that the power spectrum of the flux variability for the hot zone had peculiarities even in the absence of clear quasi-periodic oscillations in the shock location (and, consequently, the total luminosity of the hot zone). In the case under consideration, in the absence of oscillations, one should expect the power spectrum of the light curve to correspond to the power spectrum of the accretion rate at low frequencies and to be suppressed at high ones, which is explained by a finite matter cooling time in the hot zone. We numerically simulated the
problem with a given matter cooling law in the hot zone, a constant cross-sectional area of the accretion column, and a time-variable accretion rate.

\begin{figure}
\centering 
\includegraphics[width=1.1\linewidth]{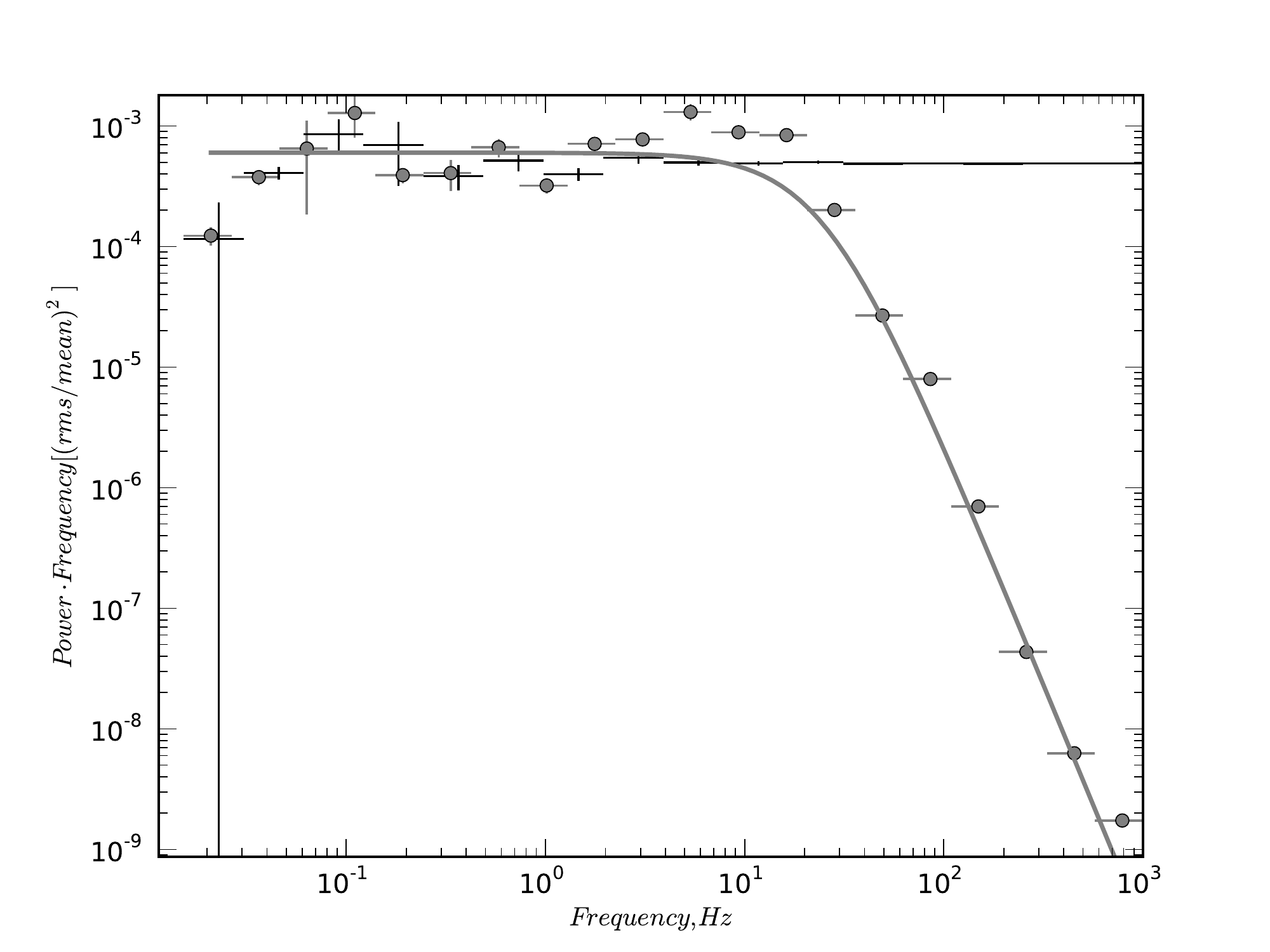}
\caption{
\small
Power spectrum of accretion rate(black cross) and X-ray luminosity from hot zone(grey circles). In presented computation combined cooling function, which contained cyclotron component $k_{cyc} \rho^{0.15} T^{2.5} $ was used. Cyclotron component was used for dumping oscillation in computation. Power spectrum was fitted by function with break, break frequency was assess as $\sim25$Hz. 
}
\label{knee}
\end{figure}

In fact, we solved a system of hydrodynamic equations similar to those considered by (\cite{langer81}).
\begin{equation}
\begin{aligned}[l]
&\frac{\partial \rho}{\partial t} = -\frac{\partial \rho v}{\partial r}\\
&\frac{\partial v}{\partial t} = - v \frac{\partial v}{\partial r} - \frac{1}{\rho}\frac{\partial P}{\partial r} - \frac{\partial \phi}{\partial r} \\
&\frac{\partial P}{\partial t} = - v \frac{\partial P}{\partial r} - \gamma P \frac{\partial v}{\partial r} - (\gamma - 1)(\Lambda_{bremss} + \Lambda_{cyc}) \\
\end{aligned}
\end{equation}
Where $P, \rho, v$ - local pressure, density and velocity. $\Lambda_{bremss}, \Lambda_{cyc}$ -are the thermal bremsstrahlung and cyclotron losses
, $\gamma$ -and is the adiabatic constant. We used the equation of state  $P = \rho \frac{k_b}{\mu m_p} T$, where $k_b$ is the Boltzmann constant.
In the problem, we assumed the cross section area of the accretion column to be constant, $A=const$, point gravity source was used $\phi = -GM/r$,  the white dwarf radius was set equal to $10^9$cm in all cases.

To numerically simulate the problem, we used the ENZO code (http://lca.ucsd.edu/portal/software/enzo/releases/
2.0), which allows 1D, 2D, and 3D problems to be simulated on an Euler grid. The PPM (Piecewise Parabolic Method) hydrodynamic simulation scheme (\cite{ppm}),which is a modified Godunov method, is implemented in this code. A quadratic interpolation is used to compute the boundary conditions between the grid cells. Second-order gravity is computed in the scheme; the acceleration due to gravity is taken into account when computing the mass fluxes. In our numerical simulations, the time step was determined as dt  $= Courant\_number \cdot dx/c$ or dt = const, where dx - is the grid spatial cell size, с - sound speed. As applied to the system of differential equations being solved, the Courant number is the
ratio of the disturbance transmission speed in a gas (the speed of sound) to the disturbance transmission
speed on a numerical grid (dx/dt). A necessary condition for the stability of the finite-difference scheme is Courant number < 1. The Courant number is a variable parameter of the problem and, as a rule, is chosen from the relationship between the accuracy
and performance requirements an excessively large Courant number (>1) will cause the scheme to be unstable; in the case of a small Courant number, it
takes more computations to obtain the solution after a time interval t. When simulating our problem, we
used two approaches: a computation with a Courant number of 0.7 (in the case of a constant accretion rate) and a computation with such a constant time that dt < dx/c (in the case of a variable accretion rate). A constant time step was used for the convenience of analysing the power spectra obtained. The energy losses are computed separately from the hydrodynamic computation. The accuracy of this computation is provided by multiple cooling computations with a dynamic cooling time step smaller than the computation time step dt. The following operations are performed in the code in one time step: a hydrodynamic computation of the change in thermodynamic parameters and velocities in the cells in time dt is performed based on the PPM scheme. After the hydrodynamic computation, the thermodynamic parameters and velocities in the cells on the grid have been updated. Subsequently, several successive matter cooling computations are made. After each cooling computation, the cell temperatures are reduced in accordance with the computed energy losses. The updated temperatures in the cells are used in each successive cooling computation. The
number of cooling computations is determined by the relationship between the total computed energy losses up to the end of the current hydrodynamic step and the current energy in the cell: the time step of the current cooling computation is  $dt_{cool} = min(dt - t_{current}, \frac{0.1 \cdot E}{dE(\rho_{current}, T_{current})/dt})$ (where $dt_{cool}$ - is the computed time to radiate all cell energy with current cooling rate , $dt - t_{current}$ - is the remaining time until the end of the computation step(= $dt - \sum{}{} dt_{cool}(previous)$), E is the current energy in the cell, dE/dt are the current computed energy losses in the cell). As a result, after the cooling computation, the grid cells have updated temperatures that are used in the succeeding hydrodynamic computation and so on. The cooling function  $\Lambda_{bremss}$ is computed by taking into account the known temperatures using the tabulated emissivities of a hot optically thin plasma with solar heavy-element abundances (\cite{Sarazin87}). To determine $\Lambda_{cyc}$ we used an analytical expression of the form  (\ref{cyc}) with the current (in each cooling computation) thermodynamic parameters in the cells. The total losses were the sum  $\Lambda_{bremss} + \Lambda_{cyc}$. 

\begin{figure}
\centering 
\includegraphics[width=1.1\linewidth]{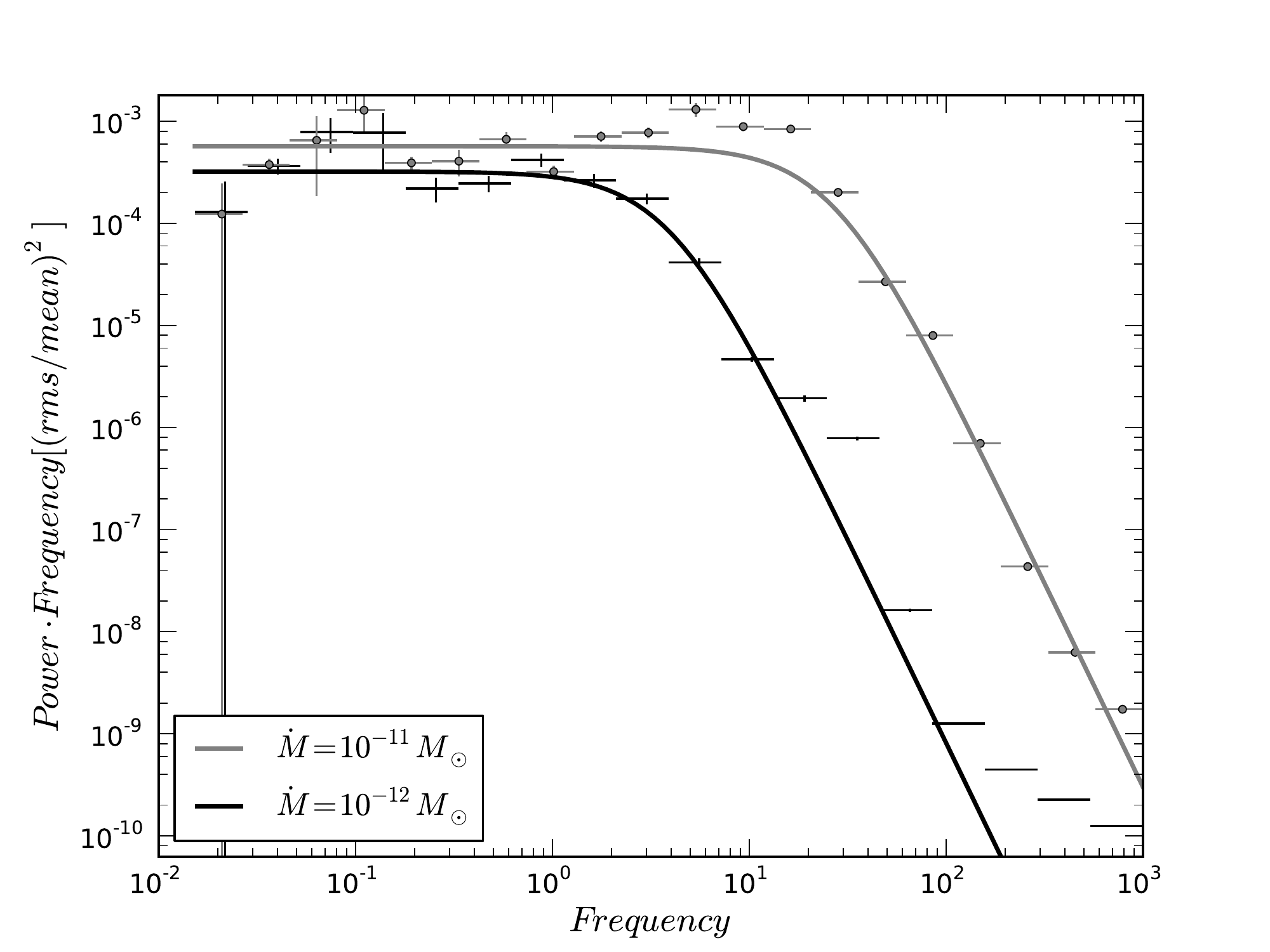}
\caption{\small 			
Power spectrums for two differ accretion rates. It's easy to see changes in break frequency due to accretion rate changes. Break frequencies is $\sim$5Hz and $\sim$25 Hz .}
\label{tar}
\end{figure}

In our computations, we used a rectangular grid with a constant spatial step, impermeable right and left walls (the accretion-column walls), and a
semipermeable lower wall (the white dwarf atmosphere). There was a gravitational field corresponding to a point source of mass $M_\odot$ at a distance of ten thousand kilometers under the lower grid wall (corresponding to the white dwarf surface) on the grid. The length from the lower grid wall to the upper one was varied, depending on the accretion rate, and was of the order of several thousand kilometers (the maximum shock height depends significantly on the accretion rate and the cooling function). The grid width from the right wall to the left one was taken to
be $10^8$cm (0.1 of the white dwarf radius) in all cases. A flow of matter with a temperature of $10^4$ K and a velocity equal to the free-fall one
 $\sim$ 5000  km sec$^{-1}$,came from the upper grid wall; the matter density depended on the current grid size and was defined as $\rho_{ff} = \frac{\dot{M}}{Av_{ff}}$. 
 
The variations in the accretion rate of matter were realized as variations in its density at a constant rate of fall and temperature. A sink of matter was realized on the lower grid wall (at the column bottom) to remove the matter cooled below a certain limit. The removal of matter is needed to avoid the "numerical collapse" of cells. Since the energy density losses in a cell are proportional to the matter density, in the case where a high density is formed in one cell, it begins to cool down faster than the neighbouring cells, even if its temperature is lower. When the
characteristic cell cooling time becomes less than the time of the hydrodynamic step, the hydrodynamic energy flux transport cannot prevent a sharp decrease in the temperature and pressure. Therefore, the case of an "ideal" absolutely impermeable wall cannot be realized in a numerical computation. The numerical collapse of a cell is characterized by a decrease in the cell temperature compared to the neighbouring cells
and by a much shorter cooling time than the time of the hydrodynamic step.

To prevent the appearance of numerical collapses on the grid during one hydrodynamic step, a series of sinks reducing the density according to the criterion $\rho' = \rho (t_{cool}/t_{hydro})$, where, $\rho'$ - is the updated density in the cell after a single sink of matter, occurs on
the lower cells. The cooling time used in computing  $\rho'$,is recalculated after each sink. As a result, by the beginning of the next hydrodynamic step, the matter density in the cell decreases to a value at which the cooling time is of the order of the hydrodynamic
step time. This sink can be identified with a phase transition when matter settles in the white dwarf atmosphere. In a numerical computation, it can give rise to additional noise on time scales of the order of the hydrodynamic step time.

In the case of bremsstrahlung losses alone and a constant accretion rate, there are the hot zone oscillations predicted by Langer (with frequencies of $\sim 1$Hz) in the scheme under consideration. As applied to our problem, the stability of the numerical scheme used is confirmed by the steady-state solution obtained in our computation in the case of combined thermal losses ($\Lambda_{bremss} + \Lambda_{cyc}$). 

\section*{Results}

Examples of the density, pressure, and velocity distributions along the accretion column derived in one of our computations for the case of cooling
through the thermal radiation of an optically thin plasma are shown in Fig.\ref{profiles}.

The light curves and height oscillations obtained in our computations in the absence of cyclotron losses are essentially similar to those from
 \cite{langer82}.  Just as Langer \cite{langer81, imamura85}  we obtained a global thermal instability for the case where the cooling of matter in the hot zone is determined by the radiation of an optically thin plasma (see Fig.\ref{qpos}).
In this computation, the accretion rate -- $10^{-11} M_\odot$/year, the accretion column area is $10^{15}$ cm, white dwarf mass - $1 M_\odot$. Accretion column height $\sim 5 \cdot 10^{8} $ cm, agrees well with the analytical estimates by  \cite{langer81}. We performed computations with a
duration of several hundred seconds in which stable oscillations were obtained for various accretion rates.

To suppress the global thermal instability resulting in quasi-periodic oscillations (which are not the goal of our paper), we introduced an additional term into the cooling function similar to the cooling through cyclotron losses of the form  $k_{cyc} \rho^{0.15} T^{2.5}$erg/cm$^3$/s. The constant $k_{cyc}$ was specified in such a way that the cyclotron losses dominated and the oscillations were suppressed. There exists some freedom of choice for $k_{cyc}$, Nevertheless, in our computations we attempted to choose the lowest values of $k_{cyc}$ for each accretion rate at which the oscillations were suppressed. 

To clearly show that the fast variability in the light curve is washed out relative to the initial variations in accretion rate, we performed the
following computation: we computed the response of the accretion-column luminosity to a short-term rise in the accretion rate (a 5\% increase in the density of infalling matter with a duration of 0.01 s Fig.  \ref{delta}). It is clearly seen that the spike in luminosity is smeared relative to the initial spike in accretion rate at the shock, implying that the high-frequency modulations in the light curve of the accretion column are suppressed.

The power spectra of the responses to signals are presented in the lower panel (Fig. \ref{delta}). To determine the break frequency, we used the analytical expression $P \propto (1 + (f/f_0)^2)^{-2}$ . The break frequency $f_0 \approx 12$ Hz is close to the estimated one (\ref{t_cooling}), which is 8 $\sim$ Hz (computation was performed for $\dot{M} = 10^{-11} M_\odot$/year;  $\Lambda_{cyc} = 2 \cdot 10^{-12} \rho^{0.15} T^{2.5}$ erg/sec/cm$^3$; $A = 10^{15}$ cm$^2; M = M_\odot; R = 10^9$cm). 

For the subsequent computations, the accretion rate was specified as an aperiodically varying quantity with a given variability power spectrum $P(f) \propto f^{-1}$. The accretion-rate variability amplitude in the frequency range  0.01-1 Hz - 5\%. 

 Figure. \ref{knee} shows the initial power spectrum of the accretion-rate variability and the power spectrum of the X-ray luminosity variability for the hot region. There is clearly seen that X-ray flux variability suppressed at high frequencies corresponding to times less then 0.1 sec. This time agrees well with the characteristic matter cooling time in the hot region for the
combined cooling law

\begin{eqnarray}
\label{t_cool_cyc}
\tau \approx \frac{\rho k_b T}{(\gamma - 1) \mu m_p (k_{bremss}  \rho^2 \sqrt{T} + k_{cyc} \rho^{0.15}T^{2.5})} = \nonumber\\
\frac{1}{(\gamma - 1)} \sqrt{\frac{3k_b}{16\mu} m_p} [ k_{bremss} (R/2GM) \dot{M}/A + \nonumber\\
k_{cyc} (\dot{M}/A)^{-0.85} (2GM/R)^{1.925} (\frac{3\mu m_p}{16k_b})^{2} ]^{-1} \nonumber\\
\end{eqnarray}

For the parameters used in numerical computation 
 ($M = M_\odot$; $A = 10^{15}$ cm$^2$ $\dot{M} = 10^{-11} M_\odot$/year; $\Lambda_{cyc} = 2 \cdot 10^{-12} \rho^{0.15} T^{2.5} $erg/sec/cm$^3$) the estimated characteristic time is 0.12 s.

 Figure. \ref{tar} shows the power spectra of the X-ray luminosity variability for the hot region in two cases where the mean accretion rate differed by a factor of 10 ($\dot{M} = 10^{-11} M_\odot/$year; (for which the estimated characteristic time is given above) and $\dot{M} = 10^{-12} M_\odot$/year; (with cyclotron loss rate $\Lambda_{cyc} = 10^{-13} \rho^{0.15} T^{2.5}$erg/sec/cm$^3$). According to  eq.\ref{t_cool_cyc}, the break frequencies of the systems are 8 and 1.5 Hz, their ratio $\sim 5$.Our estimation of the characteristic time from the above formula gives only approximate values; the ratio of the two frequencies is a more accurate parameter. Our numerical simulations yielded frequencies of $\sim$25 and $\sim$5 Hz $\omega_1/\omega_2 \sim 5$. The break frequency was estimated by using the fit  $P \cdot f \propto (1 + (f/f_0)^2)^{-2}$ ($f_0$ - is the break frequency; the function was chosen for its simplicity).

\section*{ESTIMATION OF PLASMA PARAMETERS}
	
Here, we performed numerical simulations of a supersonic plasma flow onto the surface of a white dwarf by taking into account the thermal plasma
losses through bremsstrahlung and cyclotron radiation. However, the cyclotron losses were introduced into our computations indirectly (we took the result of an approximate calculation of the energy losses in the optically thick regime of cyclotron radiation and then distributed these energy losses over the entire plasma volume below the shock; see \cite{saxton99}), therefore, it is clear that our formulas have a fairly narrow range of application. The ex act value of the energy losses was unimportant for our numerical simulations, because the goal of introducing the cyclotron losses into the plasma cooling function below the shock was to suppress the instability resulting in the oscillations of the shock location. In contrast, our main goal was the methodological question about the possibility of measuring the plasma parameters from the aperiodic accretion column flux variability in white dwarfs; therefore, the specific form of the plasma cooling function below the shock is unimportant. For this reason, to simplify our consideration of the method, let us assume that bremsstrahlung makes a major contribution to the energy losses by the hot plasma.
	
	(\cite{langer82}) showed that for the temperatures existing in the hot zone, the column is optically thin for bremsstrahlung losses. Thus, to estimate the energy losses in the accretion column, the application of the cooling function  $\Lambda_{bremss} = k_{bremss} \rho^2 \sqrt{T}$. \cite{Wu94} described an analytical solution of the steady-state problem for the height of the hot zone in the case of an impermeable lower wall, a constant column cross section, and losses of the form  $\Lambda_{bremss}$: $$h = h_0 \int_0^{\frac{\gamma - 1}{\gamma + 1}} \frac{x^2(\gamma  - (\gamma + 1) x)}{\sqrt{x(1 - x)}} dx$$, where $h_0$ - is a dimensional factor $h_0 = \sqrt{\frac{k_b}{\mu m_p}} \frac{1}{(\gamma - 1) k_{bremss}} \left(\frac{2GM}{R} \right)^{3/2}\frac{A}{\dot{M}}$ in this expression, $M$ is the mass of the white dwarf, R is its radius, $k_b$ - Boltzmann constant, $\mu$ - is the atomic weight, and x is the dimensionless velocity $x = -\frac{v}{v_{ff}}$. The characteristic time for the above problem can be estimated the same way:$$\tau = \frac{h_0}{v_{ff}} \int_0^{\frac{\gamma -1}{\gamma + 1}} \frac{x(\gamma - (\gamma + 1) x)}{\sqrt{x(1 - x)}} dx$$ (the time is assumed to be equal to the matter passage time through the hot zone). The definite integral in the above expression for the adiabatic index $\gamma = \frac{5}{3}$ is $\approx$ 0.11. It is worth noting that the relation between the column height and the characteristic
time for the adiabatic index of 5/3 in this solution turns out to be $h \approx \frac{1}{7} v_{ff} \tau$. At given white dwarf 7
mass and radius, the relation for the characteristic time depends only on the mass flux. When a white dwarf is observed, the characteristic cooling time can be directly determined from the frequency at which a break occurs in the power spectrum of the light curve. The mass flux can then be estimated from the known time:  $\frac{\dot{M}}{A} \approx 0.69/ \tau $ g/sec/cm$^2$. Thus, for example, the mass flux for a characteristic time of 0.1 s is $\frac{\dot{M}}{A}$ = 6.9 g/sec/cm$^2$. At a given X-ray luminosity (which can be estimated from the measured X-ray flux and the distance to the object), we can estimate the column area: the column luminosity $L = \frac{GM\dot{M}}{R}$, whence we obtain $\dot{M}$ and, as a result, the column area from the estimate of  $\frac{\dot{M}}{A}$ . For example for luminosity $L = 10^{32}$erg/sec the accretion rate will be $\dot{M} \approx 10^{-11} M_\odot$/year, using the mass flux and the accretion rate, we estimate the area to be: $A = 10^{14}$cm$^2$. The density in the column can be estimated from the  relation: $\rho = \frac{\gamma + 1}{\gamma - 1}\frac{\dot{M}}{A} \frac{1}{v_{ff}}$, for the mass flux obtained, the post-shock density is $5 \cdot 10^{-8}$ g/cm$^3$. The mean density in the column can be estimated from the relation $\bar{\rho} = \frac{\dot{M}}{A} \frac{\tau}{h}$ (in case of adiabatic index $\gamma = 5/3$ ~ - ~ $\bar{\rho} \approx 7 \frac{\dot{M}}{A} \frac{1}{v_{ff}}$),  in the case under consideration, the mean matter density in the hot zone is  $\bar{\rho} \approx 10^{-7}$ g/cm$^3$.

\section*{conclusions}  

We considered the temporal behavior of the luminosity of the hot region in an accretion column near the surface of a white dwarf below a standing
shock wave. We showed that the temporal luminosity variability at high frequencies ($>1$ Hz) is suppressed due to a finite matter cooling time in the hot zone($\sim 1$ sec).

We provided approximate formulas for estimating the break frequency in the power spectrum of the accretion-column luminosity variability for a given
matter cooling law. The break in the power spectrum is determined by the accretion rate, the accretion channel cross section, and the energy losses in the hot zone.

At present, the flux from white dwarfs in binary systems can be measured with a time resolution better than 0.5 - 0.1 s in both optical and X-ray spectral bands. However, since the emitting region in the optical band is comparable in size to the entire white dwarf (in the optical and ultraviolet bands, the part of the white dwarf illuminated by the accretion  column makes a major contribution to the flux) and the accretion disk, i.e., $10^9$ cm, the variability in the light curve can be washed out not only through  physical processes in the accretion column but also through a finite time it takes for the light to traverse the emitting region. This problem is not topical in the X-ray band. However, the sizes of the currently available X-ray spectrometers ( $<0.6 m^2$  ) do not allow sufficient photon statistics from such dim sources as accreting white dwarfs (with a maximum count rate of $<10$ counts/sec per 1000 cm$^2$ for the energy range 1 - 10 keV) to be accumulated. The appearance of new-generation timing spectrometers with a large collecting area can make it possible to measure such fundamental accretion-column parameters as the matter density and temperature.\\

This work was supported by the Russian Foundation for Basic Research (project nos. 10-02-00492-a, 10-02-91223-ST\_a), the Program of State Support
for Leading Scientific Schools of the Russian Federation (grant no. NSh-5069.2010.2), the Programs of the Russian Academy of Sciences P-19 and OPhN-
16, grant no. MD-1832.2011.2 from the Russian President, and State Contract no. 14.740.11.0611.\\

On the basis of V.Astakhov translation.

\end{document}